\documentclass[10pt, conference, twocolumn, compsocconf]{IEEEtran}
\usepackage{supertabular}
\usepackage{array}

\usepackage{epsfig}
\usepackage{graphicx,xspace,xcolor,cite,url}
\usepackage{times}
\usepackage{subfigure}
\usepackage{algorithmic,algorithm}
\usepackage{xspace}
\usepackage{amssymb}
\usepackage{multirow}
\usepackage{booktabs}
\usepackage{listings}
\usepackage{comment}
\usepackage{fancybox}
\usepackage{xspace}
\usepackage{paralist}

\usepackage[normalem]{ulem}
\newcommand{\xref}[1]{\S\ref{#1}}

\newcommand{\etal}[0]{\emph{et al.\xspace}\xspace}

\definecolor{groovyblue}{HTML}{0000A0}
\definecolor{groovygreen}{HTML}{008000}
\definecolor{darkgray}{rgb}{.4,.4,.4}


\lstdefinelanguage{Groovy}[]{Java}{
    keywordstyle=\color{black}\bfseries,
    stringstyle=\color{blue}\ttfamily,
    keywords=[3]{each, findAll, groupBy, collect, inject, eachWithIndex},
    morekeywords={def, as, in, use, preferences, input, definition},
    moredelim=[is][\textcolor{darkgray}]{\%\%}{\%\%},
    moredelim=[il][\textcolor{darkgray}]{§§}
}

\IEEEoverridecommandlockouts
\begin{document}

\date{}

\title{\Large \bf Internet of Things Security Research:\\ A Rehash of Old Ideas or New Intellectual Challenges?}
\author{
  \IEEEauthorblockN{Earlence Fernandes}
  \IEEEauthorblockA{University of Michigan} \and

  \IEEEauthorblockN{Amir Rahmati*\thanks{*Work done while Amir Rahmati was with the University of Michigan.}}
  \IEEEauthorblockA{Stony Brook University} \and
  
  \IEEEauthorblockN{Kevin Eykholt}
  \IEEEauthorblockA{University of Michigan} \and

  \IEEEauthorblockN{Atul Prakash}
  \IEEEauthorblockA{University of Michigan}

}

\maketitle
\thispagestyle{plain}
\pagestyle{plain}

\begin{abstract}
The Internet of Things (IoT) is a new computing paradigm that spans wearable devices, homes, hospitals, cities, transportation, and critical infrastructure. Building security into this new computing paradigm is a major technical challenge today. However, what are the security problems in IoT that we can solve using existing security principles? And, what are the new problems and challenges in this space that require new security mechanisms? This article summarizes the intellectual similarities and differences between classic information technology security research and IoT security research.
\end{abstract}

\section{Introduction}
\label{sec:intro}

Our homes, hospitals, cities, and industries are being enhanced with devices that have computational 
and networking capabilities. This emerging network of connected devices, or Internet of Things (IoT), 
promises better safety, enhanced management of patients, improved energy efficiency, and optimized 
manufacturing processes. Although there are many such benefits, security vulnerabilities in these 
systems can lead to user dissatisfaction (e.g., random bugs~\cite{barbiehack}), privacy violation (e.g., eavesdropping~\cite{iotprivacy1}), monetary loss (e.g., denial-of-service attacks~\cite{miraikrebs} or ``ransomware''~\cite{iotransom}), or even loss of life 
(e.g., attackers controlling vehicles~\cite{jeephack}). Therefore, it is critical to secure this emerging technology revolution in a timely manner. 

Although the research community has begun tackling challenges in securing the IoT, an often asked question is: What are the new intellectual challenges in the science of security when we talk about the Internet of Things, and what problems can we solve using currently known security techniques? This article summarizes some similarities and  differences between IoT security research and classic information technology security research. 

In discussing the similarities and differences, we take a broad view of the Internet of Things: we touch upon consumer-grade technologies (e.g., smart homes, smart appliances, wearables), industrial control systems (e.g., electricity grid, manufacturing), and autonomous vehicles. There are other areas of IoT such as smart cities that we consider to be outside the scope of this article. A whole set of privacy issues may arise from always-connected devices in the physical environment---this article does not go into depth on these challenges, but Davies \etal discuss possible challenges and solutions~\cite{privacyiotchasm}. Our focus is on security and safety issues.

\section{Similarities and Differences}
We classify the similarities and differences based on the standard computing stack: hardware, system software, network, and application layer. The Internet of Things computing stack is structured in a similar way: at the lowest layer we have devices that can sense and effect physical change in the environment; at the next layer we have IoT platforms that are software systems that aggregate multiple devices and controlling software to perform useful tasks; next, we have various connectivity/network protocols that enable software and physical devices to communicate with each other; and finally, we have the application layer running custom code to control physical processes. We note that it is not our aim to be exhaustive in our listing of similarities and differences.

\subsection{Hardware Layer}
The hardware layer often forms a root of trust in modern computing systems, and we expect that hardware security research results developed in the context of desktop, mobile and cloud systems to transfer in some form to IoT systems. We organize this section based on two themes: security for hardware and hardware for security.

\noindent{\textbf{Security for Hardware.}} Recent work has shown the possibility of hardware-level trojans---malicious components or instruction sequences that, when triggered, circumvent security guarantees. Yang \etal recently showed how fabrication-time attackers can inject analog components that force a flip-flop, which maintains the processor's privilege bit, to a target value~\cite{a2}. With a large percentage of IoT devices being manufactured by third-parties (often overseas), hardware-level attacks are an increasing point of concern.

\textit{Given the relative simplicity of IoT devices (e.g., sensors, microcontrollers) in comparison to general-purpose computer processors, an open question is whether such attacks can remain stealthy, and whether post-fabrication testing can be more effective in determining whether hardware trojans exist in a chip.}

\noindent{\textbf{Hardware for Security.}} Hunt \etal recently discussed ``The Seven Properties of Highly Secure Devices''---two of the properties directly concern hardware security techniques: a hardware root of trust, and hardware supported software isolation~\cite{7props}. Although, the ideas of using hardware mechanisms to securely store cryptographic keys (e.g., trusted platform modules, one-time fuses) and to create isolation units (e.g., memory management units, SGX enclaves) are similar to those in classic information technology research, we envision that many challenges will arise in \textit{applying} these notions of hardware security to IoT systems due to their limited computational and energy constraints.

These computational and energy limitations can affect higher-layer security primitives---some IoT devices may not have very precise real-time clocks, making it harder to implement even the most basic of network security protocols that assume the presence of reliable clocks. For example, Rahmati \etal showed how the natural decay rate of SRAM can be used as a time-keeper for embedded devices without clocks (e.g., smart cards)~\cite{tardis}. 

\textit{In general, we observe that although the core notions of creating hardware to support security primitives is similar to other computing paradigms, the computational and energy limitations at the hardware layer can impact security mechanisms at higher layers in the context of the IoT computing paradigm. We also observe conversely that higher-layer security properties might have to be tuned to the specific limitations of the IoT device through a hardware-software co-design approach.}

\subsection{System Software Layer}
\label{sec:softwarelayer}
The system software layer consists of firmware, operating system code, and any privileged system applications or programming frameworks. This layer builds on hardware mechanisms for establishing trust and isolation. We believe that many security principles developed in the context of mobile, desktop, and cloud computing will be applicable to IoT platforms---software systems that are similar in function to operating systems for other computing paradigms. We discuss a few areas of similarities and differences below, categorized by security principle:

\begin{itemize}
\item \textbf{Process Isolation:} This is a basic primitive that current operating systems provide---a fault in one process does not affect other processes on the system. These isolation guarantees depend on the presence of a hardware memory management unit (MMU). In small IoT devices (e.g., devices with 64KB of RAM), such an MMU is generally absent. A challenge here is to support the classic notion of process isolation without an MMU. The Tock operating system is currently exploring a combination of language-based isolation features and memory protection units (MPUs) to provide a process isolation abstraction~\cite{tock}. 

\textit{In general, although the notion of process isolation is well-known, enabling it for operating systems of resource-constrained IoT devices can require new techniques, while enabling it for IoT devices with more resources is likely not a challenge (e.g., Nest thermostats, Amazon Alexa, etc.)}

\item \textbf{Access Control:} Operating systems protect resources from untrusted code using access control. A piece of  code is either given a token (as in a capability-based system) or assigned an unforgeable unique identity upon which access control rules are expressed. Building an access control system for a particular domain is often challenging. Our prior work in analyzing consumer IoT platforms revealed access control design errors as one of the security flaws~\cite{smartthings16}. We performed an empirical security analysis of the SmartThings platform and found that access control granularity was not appropriately designed, and it led to exploitable overprivilege. A fundamental reason for such granularity design errors in access control systems stems from the tension between usability and security. This tension has manifested itself before, in mobile operating systems~\cite{feltcomphrehension}, and before them, in desktop operating systems~\cite{Bertino08someusability}. 

\textit{Although the notion of access control still applies to IoT platforms, there are new challenges in the usability aspect of designing such systems. For example, most prior access control systems dealt with virtual objects such as files and processes. In the IoT space, the objects of access control are physical devices and intuitive physical operations. An interesting challenge is exploiting our natural intuitions about physical objects while designing an access control system for IoT platforms. For example, Fernandes \etal recently discussed the notion of a user-perceived-risk-based access control system for IoT platforms~\cite{smartthings16}.}

\item \textbf{Information Flow Control (IFC):} Access control is a gatekeeper---once code obtains access to sensitive resources, access control does not provide any further protection. We analyzed a set of smart home platforms~\cite{secimplspmag}, and found that current platforms only use access control. IFC is a promising technique to control \textit{how} (untrusted) code uses its access to sensitive resources. 

\textit{Although IFC is not a new concept, as evidenced by the multitude of proposed systems for various domains, the challenge lies in applying it meaningfully to a specific domain. For example, FlowFence is a recent proposal for consumer IoT frameworks that enables a data-flow-graph approach to IFC due to the structure of IoT apps~\cite{flowfence16}. Furthermore, the kinds of confidentiality properties for environments such as homes are well-studied, however, the kinds of integrity properties that we might need, which are arguably more important in IoT, is less well-studied.}

\item \textbf{Software Updates:} Updating software is a fundamental security practice to patch security bugs, and include additional features once devices are deployed. For smartphones, personal computers, and cloud services, updating software is a well-understood, secure, and common practice. However, for physical devices in the IoT, a number of challenges arise: 
\begin{itemize}
\item Upgrading software might require a shutdown of the physical processes under control~\cite{cardenascpschallenges}, that could have economic impact.

\item Updates might require re-verification of compliance policies for safety critical devices in sensitive installations like factories and hospitals. 

\item Updates on computers in tertiary network functions (e.g., a business network) can have unintended effects on a physical process. A prominent example of a negative effect of this kind was the shutdown of a nuclear reactor due to a software update on a computer in the plant's business network~\cite{nuclearshutdown}

\item Many IoT devices deployed in the field (such as in concrete bridges) can be difficult to physically access, and might be intermittently powered (by harvesting power from vibrations). Updating the software on such intermittently powered devices is a challenge that is generally not faced in classical computing systems. 

\item IoT devices may not be updateable fundamentally because there is simply no update channel built by the manufacturers. In this case, we need to revisit our notion of a software update of the host (the device), and include notions of network-based patches~\cite{trillionunfixable}.
\end{itemize}

\textit{Although software updates for security are a well-understood concept, designing update systems for the IoT poses new challenges because of the unique properties of the physical processes that are under the control of software.}

\item \textbf{Authentication:} Passwords are currently the most widely used mechanism to authenticate users to their IoT devices, platforms, and services. But, they are also a major point of concern because weak passwords are pervasive, and have recently enabled large denial of service attacks from botnets~\cite{miraikrebs}. Although there are lightweight techniques to obtain statistical estimations of password strength,\footnote{\url{https://github.com/dropbox/zxcvbn}} weak passwords are still rampant. We do not view \textit{enforcing} reasonable strength passwords (non-default) as a technical difference from IT security, but we view it as a usability challenge. Some proposals suggest moving away from password-based authentication schemes~\cite{7props}. 

\textit{Open challenges in authenticating users to IoT devices include: (1) Are activity-based biometrics (e.g., gait, heart-rate) a better alternative to passwords given that IoT devices interact with physical phenomena? (2) IoT devices do no necessarily have classic I/O (e.g., no display in Google Home)---this can affect authentication schemes like passwords. Can we design authentication schemes of equivalent security for different interaction modalities?}

\end{itemize}



\subsection{Network Layer}
\noindent\textbf{Connectivity Protocol Diversity.} The network layer in the Internet of Things is marked by a variety of physical media and communication protocols. Part of this connectivity protocol diversity stems from the relative infancy of this technology, and part of it stems from the constraints imposed by devices or from the physical spaces that host these devices. For intermittently powered devices, short-range protocols like BLE (BlueTooth Low Energy) and NFC (Near Field Communication) are vital in conserving energy. For devices located in existing infrastructure, protocols like Physical Line Communications avoid expensive infrastructural costs. Similarly, Visible Light Communication can be useful because lights are ubiquitous in physical spaces. This protocol diversity disrupts the operation of network scanning---a fundamental security practice. We highlight this using the following case study:\\
\noindent\textit{BLE ``port'' Scanning Case Study.} In BLE, a rough analog of a TCP port is a service UUID. A device can support multiple UUIDs that define the kinds of functionality it provides. There are UUIDs for fitness machines, heart monitors etc.\footnote{See \url{https://www.bluetooth.com/specifications/gatt/services} for a list of definitions.} When a BLE device is in the disconnected state, it sends out advertisements that can help controllers (or scanners) discover the device, and attempt connections. Advertisements contain rudimentary information, and therefore, connections are required in order to get a full list of the services a device supports. Therefore, for a scanner to reliably work, a device would have to be in a disconnected state as a BLE device only accepts a single connection for its services, unlike TCP ports, where multiple simultaneous connections can be serviced on the same port. This introduces randomness into the scanning process as the scanner will have to ``try again'' at a later point in time in the hope that the BLE device is in the disconnected state. Furthermore, if a BLE device is in the connected state, it does not send advertisements, further complicating scanner operation.\footnote{Sophisticated scanners could try to jam existing connections to force them to drop.}

Therefore, scanners for IoT protocols are currently very network-specific and only offer limited coverage (BLE scanners will only be useful for BLE devices but it is common for physical spaces such as a home to contain devices using different connectivity protocols). This is in stark contrast to the Internet in general where TCP/IP is a constant presence for online services where network scanning is generally used. Port scanning is further made difficult in the consumer IoT space due to the practice of placing devices behind a hub or router. Network scanners situated outside such a network will not be able to conduct internal scans.

\textit{As each protocol has its own notions of how two peers communicate with each other, it is unclear how network security practices such as port scanning translate to networks of devices that use various IoT protocols.}

\noindent\textbf{Re-purposing of Networking Technologies in Unforeseen Ways.} As discussed above, a common IoT system architecture for smart homes is to connect multiple devices to a hub. If all the home IoT devices use WiFi as a connectivity protocol, then a WiFi router can be a hub. This kind of configuration poses new security challenges that WiFi was not designed to support. For example, it is very difficult to ensure that only a WiFi-enabled presence detector should affect a door lock. Such an isolation boundary is useful because there could be multiple devices on a network, some of which might be malicious or compromised through bugs. The isolation unit would serve as defense in depth against such a situation. Furthermore, as discussed in~\xref{sec:softwarelayer}, some devices may not have update channels, necessitating other means of updates. A central hub like a WiFi router can be in a good position to apply updates in the form of filters for known malicious traffic patterns. Simpson \etal discuss the design of a WiFi home hub that can perform such security functions~\cite{securehub}. 

\textit{In the context of smart homes, we observe that hubs like WiFi routers are being increasingly used to support IoT device networks. Adapting these hubs to support security properties such as isolation as first-class citizens is an open challenge.}

\noindent\textbf{Anomaly Detection in the Network.} As defense in depth, detecting misbehaving devices on the network is a common and well-deployed security practice in many computing areas. The main challenge in obtaining useful results from anomaly detectors is tuning it to produce a low number of errors---either raising a flag for benign behavior or not raising a flag for malicious behavior. This challenge arises due to the fundamental complexity of the devices we typically connect to a network---general purpose computers like mobile phones, desktops, and servers. These devices perform multiple functions, and lead to complicated network traces that make it difficult to characterize ``normal'' behavior. In contrast, IoT devices are simple and have a single purpose. This can translate to simpler network dynamics, and hence easier to model behaviors ultimately leading to a lower number of errors in anomaly detectors. Recent work in the context of industrial control systems show promising results---Formby \etal show how predictable network characteristics of relays and circuit breakers can be used to reliably fingerprint them~\cite{formbyndss16}.

\textit{A physical process evolves as per physical laws in a generally predictable fashion. A garage door of a certain mass takes a specific amount of time to close, and an oven of a certain volume would heat up to a specific temperature in a predictable amount of time. We envision that models of these physical processes can be used to reduce the errors in anomaly detectors. In contrast, general purpose computers, by definition, do not have well-defined models of behavior when applications running on them are taken into account.}

\subsection{Application Layer}
\label{sec:applayer}
The application layer in IoT is no different from other computing paradigms---it runs customized code for end-user scenarios. In this section, we consider two ways in which IoT application behavior can affect security.

\noindent\textbf{Physical Co-Relations.} Consider a simple If-This-Then-That rule that closes a garage door after 9PM. If a speaker were placed in the vicinity of the motors controlling the door, it would record a specific acoustic pattern for a specific amount of time whenever the door closes. There is a natural physical co-relation between this acoustic pattern and the closing of the garage doors. 

\textit{The natural co-relations between physical phenomena could act as feedback channels that IoT platforms can use to approximately monitor physical processes for deviations from expected behavior. If deviations occur, then it could imply a failure or a security issue.}

\noindent\textbf{Machine Learning and Control of Physical Processes.} In recent years, machine learning (and deep learning) has found wide applicability to many domains of computing---deep learning robots can learn to grasp objects, and the Nest thermostat can learn and then control HVAC settings automatically. However, recent work has shown that deep learning algorithms are susceptible to adversarial manipulations of their inputs---attackers can craft inputs that look indistinguishable from benign inputs to humans, but can be interpreted in a completely different way by machines. For example, tampered images that are fed into a vision algorithm running on an autonomous vehicle can make the vehicle believe a stop sign was a yield sign, causing a possible crash at an intersection. Building robustness into ML algorithms against such attacks an active area of research whose details are beyond the scope of this article. We refer the reader to~\cite{DBLP:journals/corr/PapernotMSW16} for a more thorough treatment of the topic.

\textit{As more physical processes come under the control of machine learning algorithms, their vulnerabilities in adversarial settings will become pressing security and safety issues. Classic IT security has often applied ML to security problems (e.g., malware detection), however, only recently has work begun on securing the ML algorithms themselves.}

\section{Concluding Thoughts}
Broadly, the similarities between classic IT security research and IoT security research are the basic secure software and hardware construction principles that have been developed in other computing paradigms. The differences form a spectrum of new intellectual challenges. On one end of this spectrum, challenges arise in applying and adapting known security principles to make them work for the unique challenges posed by the IoT computing paradigm. We believe that overcoming many of these challenges will involve a cross-layer co-design approach. For example, due to limited energy availability, hardware security mechanisms might need to be purpose-built depending on the specific higher-level security property we wish to enforce---it is not possible to efficiently accommodate a one-size-fits-all security mechanism.

At the other end of the spectrum, the nature of physical processes and the nature of IoT devices lend themselves to the construction of new security mechanisms. As discussed, natural co-relations between physical phenomena can be exploited to detect security and safety failures. Similarly, the predictability of physical processes is another avenue that can be used to detect anomalous events. Finally, introducing ideas from the control engineering world into IoT platform construction (e.g., specialized feedback loops) could lead to a more secure and safe IoT.






\bibliographystyle{IEEEtranS}
\bibliography{references}

\end{document}